\documentclass[aip,reprint]{revtex4-1}
\usepackage{graphicx}
\usepackage{epstopdf}
\usepackage{booktabs}
\usepackage{grffile}

\draft % marks overfull lines with a black rule on the right

\begin{document}

%\preprint{}

\title{Dosimetric and Deformation Effects of Image-Guided Interventions during Stereotactic Body Radiation Therapy of the Prostate using an Endorectal Balloon} %Title of paper



\author{Bernard L. Jones}
%\thanks{}
\altaffiliation{Author to whom correspondence should be addressed: bernard.jones@ucdenver.edu}
\affiliation{Department of Radiation Oncology, University of Colorado School of Medicine}
\author{Gregory Gan}
\affiliation{Department of Radiation Oncology, University of Colorado School of Medicine}
\author{Quentin Diot}
\affiliation{Department of Radiation Oncology, University of Colorado School of Medicine}
\author{Brian Kavanagh}
\affiliation{Department of Radiation Oncology, University of Colorado School of Medicine}
\author{Robert D. Timmerman}
\affiliation{Department of Radiation Oncology, University of Texas Southwestern Medical Center}
\author{Moyed M. Miften}
\affiliation{Department of Radiation Oncology, University of Colorado School of Medicine}


%\date{\today}
\newcommand{\super}[1]{\ensuremath{^{\textrm{{\tiny #1}}}}}
\newcommand{\subsc}[1]{\ensuremath{_{\textrm{{\tiny #1}}}}}
\newcommand{\enm}[1]{\ensuremath{#1}}

\begin{abstract}
\textbf{Purpose:} During Stereotactic Body Radiotherapy (SBRT) for the treatment of prostate cancer, an inflatable endorectal balloon (ERB) may be used to reduce motion of the target and reduce the dose to the posterior rectal wall. This work assessed the dosimetric impact of manual interventions on endorectal balloon position in patients receiving prostate SBRT, and investigated the impact of ERB interventions on prostate shape.

\textbf{Methods}: The data of seven consecutive patients receiving SBRT for the treatment of clinical stage T1cN0M0 prostate cancer enrolled in a multi-institutional, IRB-approved trial were analyzed. The SBRT dose was 50 Gy in 5 fractions to a planning target volume (PTV) that included the prostate (implanted with three fiducial markers) with a 3 mm margin. All plans were based on simulation images that included an ERB inflated with 60 cm\super{3} of air.  Daily kilovoltage (kV) cone-beam computed tomography (CBCT) imaging was performed to localize the PTV, and an automated fusion with the planning images yielded displacements required for PTV re-localization. When the ERB volume and/or position were judged to yield inaccurate repositioning, manual adjustment (ERB re-inflation and/or repositioning) was performed. Based on all 59 CBCT image sets acquired, a deformable registration algorithm was used to determine the dose received by, displacement of, and deformation of the prostate, bladder, and anterior rectal wall.  This dose tracking methodology was applied to images taken before and after manual adjustment of the ERB (intervention), and the delivered dose was compared to that which would have been delivered in the absence of intervention.

\textbf{Results}: Interventions occurred in 24 out of 35 (69\%) of the treated fractions.  The direct effect of these interventions was a significant increase in the prostate radiation dose that included 95\% of the PTV (D95) from 9.6 \enm{\pm} 1.0 Gy to 10.0 \enm{\pm} 0.2 Gy (p=0.06) and a significant increase in prostate coverage from 94.0 \enm{\pm} 8.5\% to 97.8 \enm{\pm} 1.9\% (p=0.03).  Additionally, ERB interventions reduced prostate deformation in the anterior-posterior (AP) direction, reduced errors in the tilt of the prostate, and increased the similarity in shape of the prostate to the radiotherapy plan (increased Dice coefficient from 0.76 \enm{\pm} 0.06 to 0.80 \enm{\pm} 0.04, p=0.01).  Post-intervention decreases in prostate volume receiving less than the prescribed dose and decreases in the voxel-wise displacement of the prostate, bladder, and anterior rectal wall were observed, which resulted in improved dose volume histogram (DVH) characteristics.  

\textbf{Conclusions}: Image-guided interventions in ERB volume and/or position during prostate SBRT were necessary to ensure the delivery of the dose distribution as planned.  ERB interventions resulted in reductions in prostate deformations that would have prevented accurate localization of patient anatomy. 

\end{abstract}


\maketitle %\maketitle must follow title, authors, abstract and \pacs

%\section{}
%\label{}
%\subsection{}
%\subsubsection{}

% If in two-column mode, this environment will change to single-column format so that long equations can be displayed. 
%\begin{widetext}
%$$\mbox{put long equation here}$$
%\end{widetext}

% Figures should be put into the text as floats. 
% Use the graphics or graphicx packages (distributed with LaTeX2e).
% Put the label that you will use with \ref{} command in the braces of the \label{} command.
%

\section{Introduction}

As external beam radiation delivery techniques become more sophisticated, there has been a trend towards using fewer fractions with a higher dose, or “hypofractionation\super{1}.” An example of this is a common treatment known as Stereotactic Body Radiation Therapy (SBRT), in which an entire course of radiotherapy is given in only five (or fewer) fractions.  These treatments can be advantageous due to radiobiological concerns, therapeutic ratio, and patient convenience.  However, care must be taken to ensure accurate treatment planning and delivery, as these hypofractionated regimens usually involve very high dose per fraction and relatively tight expansion margins around the target volume.

SBRT can provide excellent clinical outcomes for selected patients with prostate cancer\super{2-7}, a result thought to stem from prostate cancer having a relatively low \ensuremath{\alpha}\/\ensuremath{\beta} ratio, which favors a hypofractionated approach\super{8,9}.  Critical to the success of SBRT are careful efforts to limit the volume of the rectum and bladder that receive a high dose.  The prostate surrounds the urethra as it exits the bladder and sits directly anterior to the rectum, making it difficult to avoid these structures.  Another complicating factor is the contents of these two organs, as changes in urine/stool volume can displace and deform the prostate\super{10-15}, making it difficult to localize using rigid transformations of the patient anatomy.

 \begin{figure*}
 \includegraphics{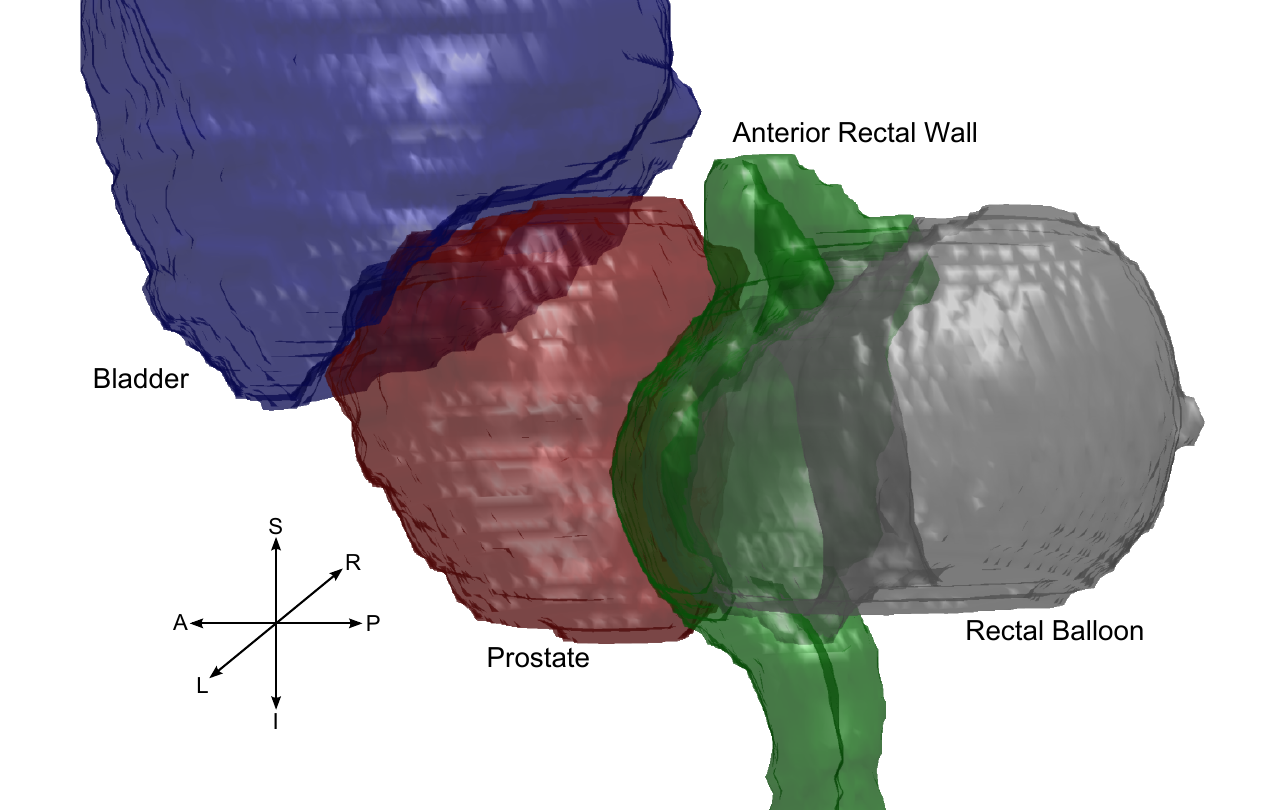}%
 \caption{3D geometry of prostate treatment.  The prostate lies inferior/posterior to the bladder and anterior to the rectal wall.  The ERB expands the rectal wall and exerts force against the prostate in the anterior direction.  The organs-at-risk (OAR) for this treatment include the bladder and anterior rectal wall.}%
 \end{figure*}

In many cases, an inflatable endorectal balloon (ERB) is used during hypofractionated prostate treatment.  The ERB serves to reduce intrafraction motion of the target\super{16-18}; however, care must be taken when actively displacing the prostate via ERB use, especially in SBRT when margins are small and dose per fraction is high.  Many studies have observed that in the similar situation of inflatable endorectal coil use for MRI imaging of the prostate, the gland may be compressed by as much as 1 cm in the anterior-posterior (AP) direction\super{19-23}.  Proper use of the ERB can also decrease rectal toxicity, as it displaces the posterior rectal walls out of the high dose regions\super{24-26}.  Apart from a slight skin sparing effect at the air/wall interface\super{27}, the ERB does not significantly affect the dose absorbed by the anterior rectal wall\super{24}, and thus when employing an ERB the rectal toxicity seen in prostate radiotherapy\super{28-30} is dependent on the anterior rectal dose.  With these factors in mind, it is necessary in many cases to manually adjust the ERB volume and/or position (intervene) after initial patient set-up to allow for maximum agreement of the PTV, bladder, and rectal wall positions between the radiotherapy plan and treatment positions.

A similar situation exists in conventional radiotherapy treatment of the prostate bed.  An ERB is not commonly used; however, changes in urine/stool volume can displace and deform the prostate.  A previous study examined the dosimetric impact of manual interventions on rectal and bladder filling (i.e. stool and/or urine volume) in this scenario\super{31}.  This study concluded that the low frequency of interventions (16\%) and the small prescription dose per fraction (2 Gy) in a conventional fractionation scheme diluted the effect of manual changes in rectal and bladder filling, and there were no significant differences in overall outcome from interventions when daily cone beam computed tomography (CBCT) localization was performed.  However, because the conditions are different for SBRT with ERB use, the effect of interventions could be significant.

In this work, the dosimetric impact of manual interventions of the ERB volume and/or position in patients receiving SBRT to the prostate was assessed.  This was accomplished by retrospectively analyzing daily CBCT images taken before and after manual adjustment of the ERB (intervention), and comparing the delivered dose to that which would have been delivered in the absence of intervention.  This work also assessed the relationship between ERB interventions and deformation of the shape of the prostate.

\section{Methods}

\subsection{Patients}

The data (59 CBCT image sets) of seven consecutive patients with clinical stage T1cN0M0 prostate cancer who were enrolled in our institution as part of a multi-institutional, IRB-approved trial of SBRT for low-to-intermediate risk prostate cancer were studied\super{32}.  The patients (median age: 64 years; range: 56-69 years) were treated with either static beam intensity modulated radiotherapy (IMRT, n=1) or volumetric modulated arc therapy (VMAT, n=6) on an Elekta Synergy accelerator (Elekta, Crawley, UK) between January and December 2010.  

 \begin{figure}[h]
 \includegraphics{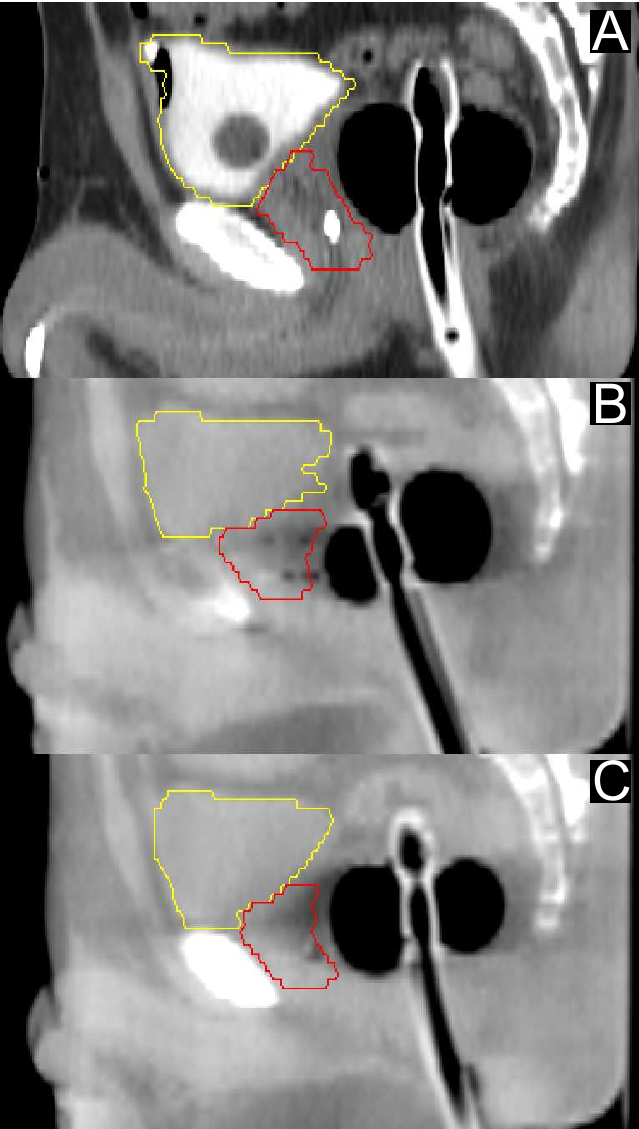}%
 \caption{Saggital slices of the pelvis showing the prostate (red), bladder (yellow), and effect of ERB intervention.  A) pCT image showing radiotherapy planning position.  B) CBCT image showing pre-intervention ERB position.  The ERB was not inserted to the same depth as it was for the pCT scan, and also exhibits a relative tilt.  The shape of the prostate is noticeably different.  C) CBCT image showing post-intervention ERB position, which more closely matches the pCT image.}%
 \end{figure}

\subsection{Simulation, planning, and delivery}
Prior to simulation, fiducial markers were implanted into the prostate to assist in localization. To stabilize the prostate and displace the posterior rectal wall, an ERB inflated with 60 cm\super{3} of air was used both during simulation and delivery.  Patients were simulated and treated in the supine position.  Patients were simulated on a CT scanner utilizing contrast to identify the bladder and urethra.  Treatment planning was performed using either the XiO IMRT or Monaco VMAT treatment planning systems (Elekta, St. Louis, MO).  The PTV was generated using expansion margins of 3 mm in all directions from the prostate, and was prescribed to receive 50 Gy in five fractions to 95\% of the volume. 

Per the clinical trial protocol\super{32} the following structures were contoured, and the following dose limits were observed.  The rectal wall was contoured from the superior edge of the anal sphincter (inferiorly) to 1 cm above the superior extent of the prostate.  Assigning 0º to the most anterior aspect of the rectum at mid-sagittal plane, the anterior rectal wall was contoured from 315º to 45º.  The contents of the bladder were excluded from the bladder contour.  The maximum point dose in the anterior rectal wall, bladder, and prostatic urethra was limited to 105\% of the prescription dose.  Additionally, no more than 10 cm\super{3} of the bladder could receive in excess of 18.3 Gy.  All dose constrains were followed.  Since the clinical trial specifies a small PTV expansion margin (3-5 mm), accurate delivery requires immobilization using an ERB, localization using fiducial markers, daily image guidance using CBCT, and physician review/approval for each delivered fraction. Full description of the protocol, patient eligibility, treatment planning and dose constraints are provided elsewhere\super{32}.

Before each fraction, localization of the prostate was accomplished by acquiring a kilovoltage (kV) CBCT scan and fusing it to the planning CT (pCT).  By applying rigid body translations to the CBCT, shifts were determined which resulted in the closest match between the two scan positions.  The first step in localization was to align the three fiducial markers implanted in the prostate.  Second, the alignment of the ERB was compared to the pCT.  In 31\% of the treatments, internal anatomy on the CBCT and pCT matched sufficiently well to proceed with treatment without manual adjustment (intervention).  

For 24 of the 35 treatment fractions (69\%), the initial ERB placement was clinically judged to be insufficient during localization to allow maximization of PTV coverage and organs-at-risk (OAR) sparing using only rigid-body translations.  For instance, it was seen in some cases that the CBCT prostate extended outside the superimposed PTV contour. In other cases, misalignment of the anterior ERB edge after alignment of the fiducial markers was observed. For these cases, the position of the ERB was manually re-adjusted to achieve better alignment.  An overview of the target and OAR geometry is shown in Fig. 1, and an example ERB intervention is shown in Fig. 2. In total, there were 24 pre-intervention image sets, 24 post-intervention sets, and 11 fractions where no intervention occurred. The act of intervening added roughly 5-15 minutes to the total patient setup time.

\subsection{Deformable Registration}
To assess interfraction spatial displacement and dose changes to individual voxels, the prostate, bladder, and anterior rectal wall were contoured on each pre- and post-intervention CBCT (59 total CBCT image sets).  The contours were drawn by the same physician and reviewed by the same physicist for intra- and inter-observer consistency.  The voxel tracking process was adapted from a previous study of prostate bed deformations\super{31} and proceeded as follows using software developed by the current authors.  First, a set of matching control points for each organ was computed based on the CBCT and pCT contours.  The control points were selected in order to best capture the known mechanics of deformation in the prostate and rectal wall.  It has been shown that ERBs lead to little prostate deformation along the superior-inferior axis\super{20}. It is also generally assumed that changes in rectal filling only stretch the rectal wall transversely\super{33}, and this assumption is reinforced by the constraints placed on the rectal wall contour in the axial direction\super{32} (i.e. uniform axial length).  Thus, the control points for each slice were selected along the contours at equal angular intervals from the slice centroid, which captures expansion and contraction of the contours in the axial plane while preventing non-physical rotation that can occur in deformable registration of spheroid/cylindrical shapes, such as the prostate, bladder, and rectum.

Based on the displacements between matching control points, a pCT-to-CBCT deformation field was generated for each organ using thin-plate spline warping\super{34}.  Using this deformation field, each voxel within the pCT contour was deformed to a new location in the CBCT.   To calculate the pre-/post-/no intervention voxel-wise displacement of each organ, couch shifts were applied to the CBCT dataset such that the coordinates of the scan represented the geometry that was either used for treatment (in the case of post- or no intervention) or would have been used for treatment in the absence of intervention (pre-intervention).  

 \begin{figure}[h]
 \includegraphics{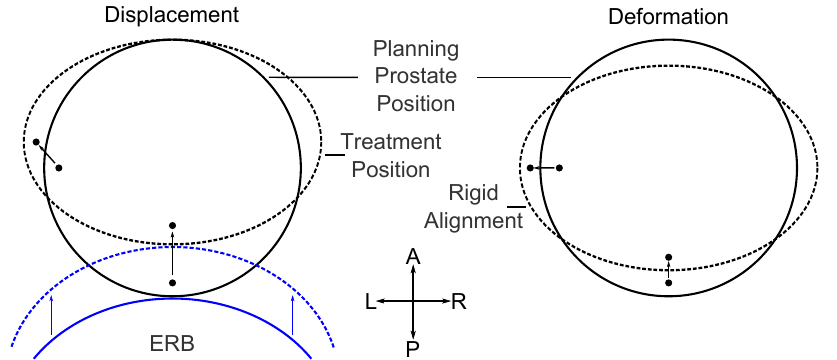}%
 \caption{Axial slice of the prostate and ERB demonstrating displacement and deformation.  Left) The ERB is placed incorrectly at time of treatment, causing the prostate to shift in the AP direction from the plan position.  A deformable algorithm is used to determine the displacement of each voxel in the prostate in order to determine dose received during treatment.  Right) The same algorithm is applied after rigid alignment of the two prostate structures in order to quantify the deformation (i.e. difference in shape) of each voxel.}%
 \end{figure}

\subsection{Displacement/Deformation and Dose}
Using the pCT-to-CBCT deformation field, the vector translation of each voxel from its position in the pCT geometry to a new position in the CBCT geometry was calculated.  By applying this methodology to each CBCT image set, the location of each voxel in the pCT structures was tracked throughout all pre-/post-/no intervention treatments.  Doses were not recalculated on the CBCT images due to effects such as x-ray scatter, detector glare, and image lag, which cause the images to contain CT-to-electron density variations too large for accurate dose calculation\super{35}. To calculate the dose received in the pre- or post-intervention geometry, each voxel was translated within the planning dose field (based on the pCT-to-CBCT deformation field).  In other words, the planning dose field was superimposed on the CBCT geometry. 

Based on the displacement and dose information, several dosimetric parameters were calculated for pre- and post-intervention treatments.  \textit{PTV Coverage} was defined as the fraction of the deformed CBCT prostate volume that was within the PTV.  \textit{Dx} was the lowest dose received by the highest x\% of the volume, and \textit{Vx} was the volume receiving \enm{>}x dose.

To analyze the deformations in prostate shape, a second pCT-to-CBCT deformation field was calculated.  First, a rigid registration was performed between the pCT and CBCT prostate contours by minimizing the total distance between matching control points.  Then, using the previously described methodology, the vector translation of each voxel in the prostate from the pCT to CBCT was calculated.  By aligning the two structures before applying the control point-based deformation field, the calculated deformations more accurately represent changes in the shape of the prostate, and exclude any systematic shifts that may be present due to imperfect alignment or prioritization of OAR sparing above target coverage.  Throughout the following text, “deformation” refers to the changes in shape (i.e. vector translation of each voxel based on pCT-to-CBCT deformation field after rigid alignment of the prostate), while “displacement” refers to changes in absolute position, or in other words changes in shape in addition to any rigid-body misalignment.  The differences between these two quantities are demonstrated in Fig. 3.

 \begin{figure*}
 \includegraphics{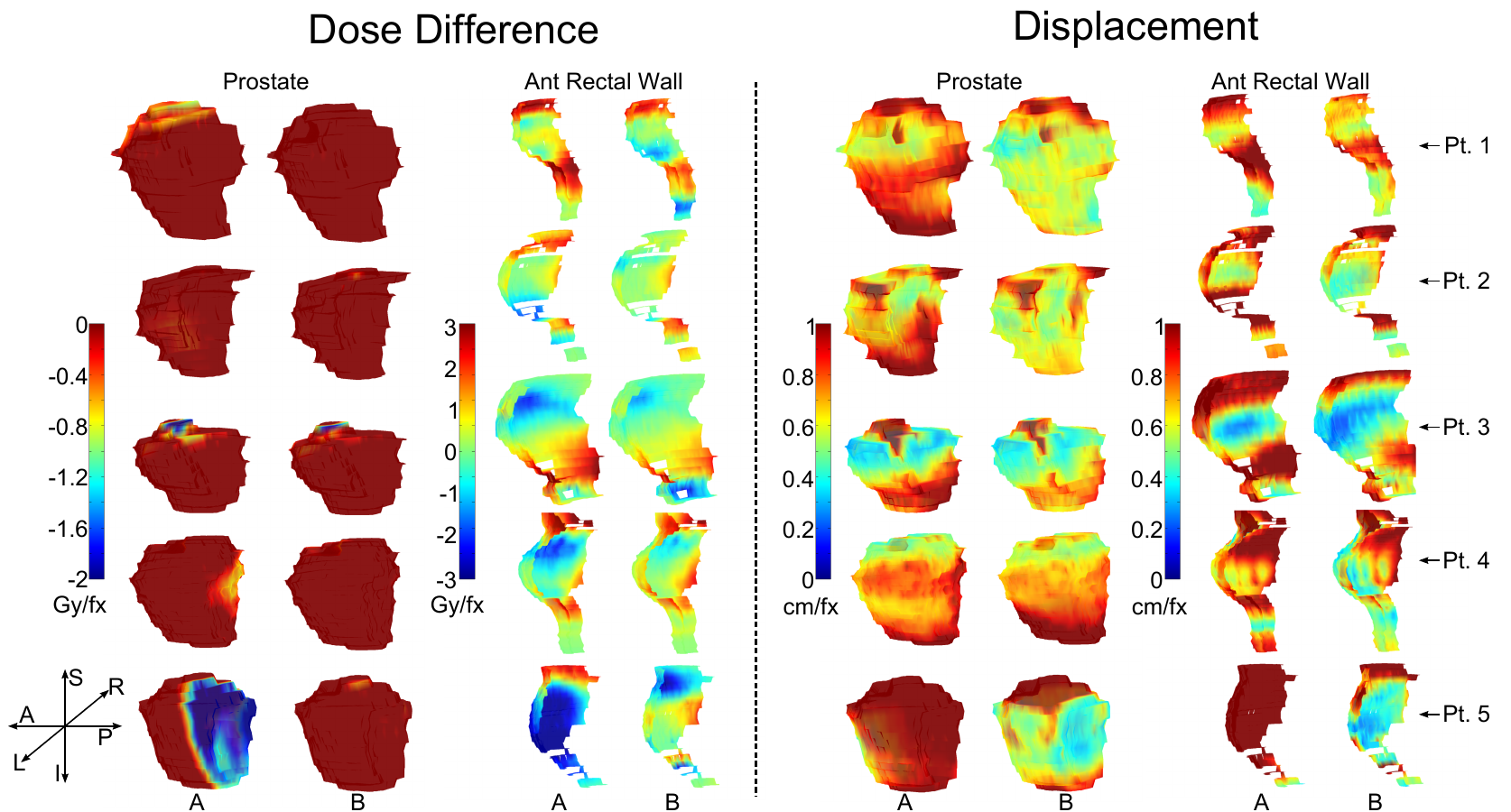}%
 \caption{Patient-specific changes in dose to and displacement of the prostate and ARW, where each row represents the prostate/ARW of a different patient.  For each patient with \ensuremath{>} 1 intervention, the dose and displacement of each voxel was averaged within all pre- and post-intervention treatments (denoted as columns \textit{A} and \textit{B}, respectively), based on the voxel-wise displacements computed from CBCT contours.  Doses on the surface of the prostate are displayed as the difference from the prescription dose of 10 Gy/fx; thus, any non-zero data points represent under-dosing of the prostate. Doses to the anterior rectal wall are displayed as the average difference per fraction from the treatment plan dose.  In the prostate, the cumulative effect of intervening was an increase in target coverage and a decrease in surface displacement from the planning prostate shape.  In the ARW, the cumulative effect of intervention was that the resultant doses more closely resembled the plan dose, and that there was a decrease in surface displacement near the ERB/prostate interface. }%
 \end{figure*}

To quantify differences in the shape of the prostate, the Dice coefficient was calculated between each pCT and CBCT prostate contour (after rigid alignment).  The Dice coefficient (\textit{D}) is a measure of the similarity of two pixel sets, and ranges from 0 to 1 for shapes with no overlap to perfect overlap, respectively.  The Dice coefficient\super{36} is calculated via Eq. 1, where \enm{|C_1{\cap}C_2|} is the volume of all overlapping voxels between two contours, and \enm{|C_n|} is the total volume of a contour.

\begin{equation}
D=\frac{2|C_1{\cap}C_2|}{|C_1|+|C_2|}
\end{equation}

\section{Results}
\subsection{Single-fraction Impact}
Since the SBRT technique involves a higher dose per fraction and a smaller number of fractions (compared to conventional treatments), significant deviations in a single fraction could have a large impact on the overall treatment outcomes.  In order to compare the effects of ERB adjustments between patients, dosimetric parameters of interest were calculated for each fraction in which an ERB adjustment (intervention) was performed (Fig 2).  The direct dosimetric effects of ERB adjustments are shown in Table 1 for PTV, anterior rectal wall (ARW), and bladder (BLA).  These data are the mean values based on the paired results of all 48 pre- and post-intervention CBCTs.  P-values are shown for a paired t-test of the hypothesis of equal means. 

\begin{table}[htbp]
  \centering
  \caption{Single-fraction dosimetric effect of manual ERB adjustment (n=24)}
    \begin{tabular}{rccccc}
    \hline
    \textbf{} & \multicolumn{2}{c}{Pre-Intervention} & \multicolumn{2}{c}{Post-Intervention} & p \\
    \hline
    \multicolumn{1}{c}{PTV D95 (Gy)} & 9.64  & \enm{\pm}1.0  & 10    & \enm{\pm}0.2  & 0.06 \\
    \multicolumn{1}{c}{PTV Cov (\%)} & 94.6  & \enm{\pm}7.6  & 98.0    & \enm{\pm}1.9  & 0.03 \\
    \multicolumn{1}{c}{ARW V6 (cm\super{3})} & 9.1   & \enm{\pm}2.8  & 9.3   & \enm{\pm}2.2  & 0.47 \\
    \multicolumn{1}{c}{ARW V8 (cm\super{3})} & 7.1   & \enm{\pm}3.0  & 7.4   & \enm{\pm}2.4  & 0.49 \\
    \multicolumn{1}{c}{ARW V10 (cm\super{3})} & 3.3   & \enm{\pm}1.6  & 2.7   & \enm{\pm}1.2  & 0.17 \\
    \multicolumn{1}{c}{BLA V2 (cm\super{3})} & 31    & \enm{\pm}12   & 29    & \enm{\pm}10   & 0.17 \\
    \multicolumn{1}{c}{BLA V4 (cm\super{3})} & 20    & \enm{\pm}11   & 19    & \enm{\pm}8.9  & 0.45 \\
    \multicolumn{1}{c}{BLA V6 (cm\super{3})} & 14    & \enm{\pm}10   & 13    & \enm{\pm}7.8  & 0.36 \\
    \hline
    \end{tabular}%
	\enm{} Abbreviations: PTV=planning target volume, Cov=coverage, ARW=anterior rectal wall, BLA=bladder, Vx=volume receiving \enm{>}x Gy per fraction, Dx=lowest dose received by the highest x\% of the volume.
  \label{tab:addlabel}%
\end{table}%

It can be seen that there were significant increases in both D95 to the prostate and in the prostate coverage.  However, there was no significant change in the dosimetric parameters of the anterior rectal wall or bladder.  This was due to the fact that the primary objective in aligning the patient was target coverage, thus interventions may either increase or decrease dose to normal tissue. 

 \begin{figure*}
 \includegraphics{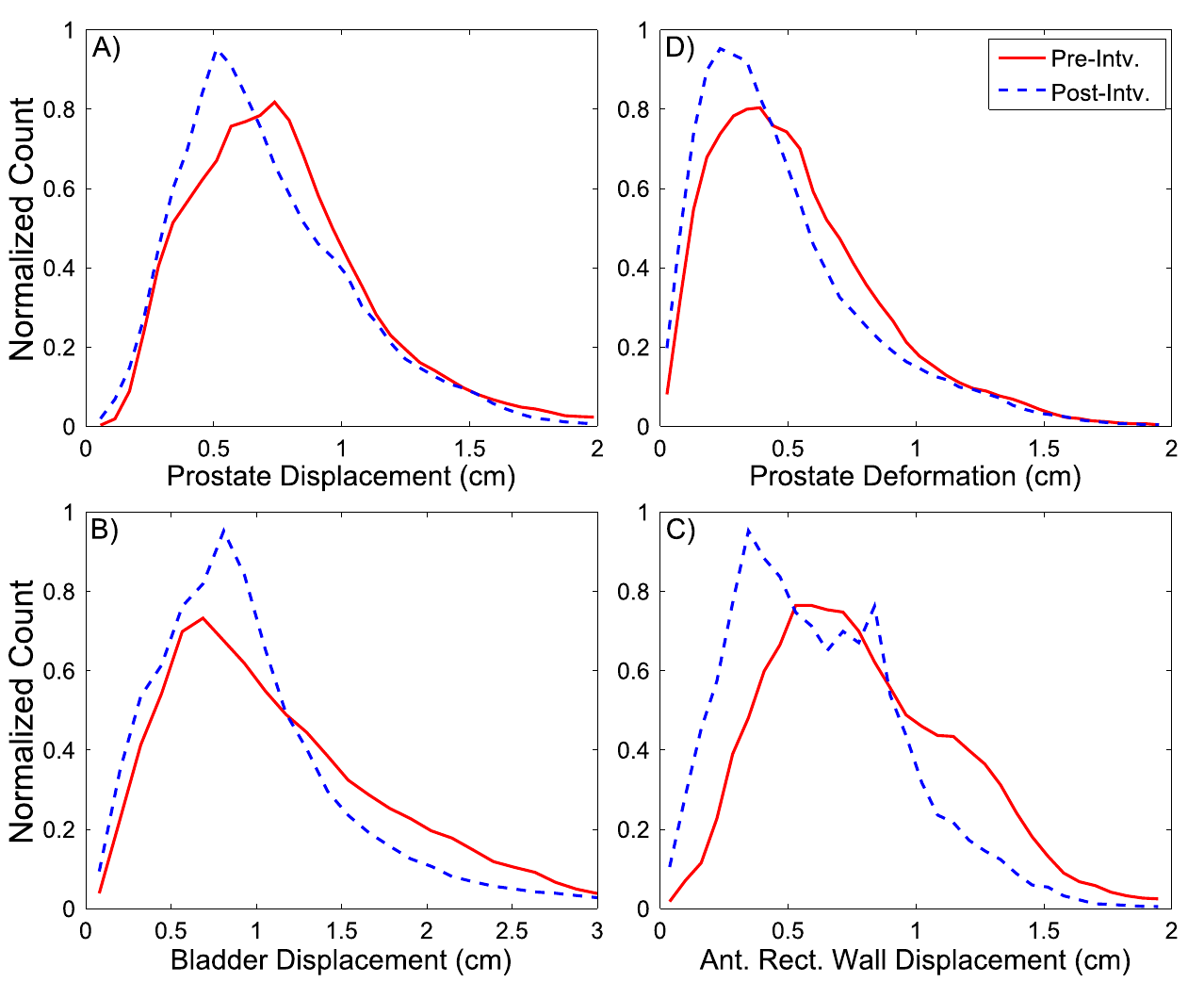}%
 \caption{Effect of ERB interventions on the position of each voxel in the prostate, anterior rectal wall, and bladder. “Deformation” represents the changes in shape, while “displacement” represents changes in shape in addition to any rigid-body misalignment.  In all cases (a-c), the displacement of each organ relative to the radiotherapy plan was reduced as a result of interventions.  Additionally, the deformation of the prostate (d) was reduced.}%
 \end{figure*}

\subsection{Patient-specific Impact}
Interventions were performed for 69\% of all treated fractions, with an average of three interventions per patient.  To visualize the effects that interventions could have on a single patient, Fig. 4 shows the dose difference and displacement along the surface of the prostate and ARW for all patients with more than 1 intervention.  Each row in these figures shows the prostate/ARW of a different patient.  For the prostate, the dose is displayed in terms of the difference from the planned dose of 10 Gy/fraction; thus, any non-zero data points represent under-dosing of the prostate.  For the ARW, the dose is displayed as the difference from the radiotherapy plan dose for that voxel.  In both cases, the displacement is shown in terms of average absolute displacement of each voxel from the pCT (based on the pCT-to-CBCT deformation field).  To allow comparison between patients with different numbers of interventions, each plot shows the dose and displacement changes averaged over all intervened fractions (mean change per intervened fraction).

The number of voxels in the prostate receiving less than the prescription dose was decreased as a result of interventions.  This can be attributed to the reductions in displacement also seen in Fig. 4.  A particularly large change is seen in the last row of Fig. 4; in that case, the ERB did not inflate properly, leading to very large deformations in the shape of the prostate and rectum. In the ARW, note that these values can yield insight into the ERB position/ARW shape during treatment.  For instance, a dose increase on the ARW surface below the visible ERB bulge implies that the ERB was slightly lower during treatment, pushing this inferior portion of the ARW into the treatment field.  In nearly all voxels, the cumulative effect of intervention was that the resultant dose distributions more closely resembled the plan dose, especially along the ARW/prostate interface, which receives the highest dose.  The displacement seen in this area was also greatly reduced. 

\subsection{Cumulative Impact}
To understand the effect of ERB interventions on the position and shape of the prostate, ARW, and bladder, the displacement of each voxel for all 24 pre- and post-intervention fractions was pooled and analyzed.  Histograms of the absolute displacement per voxel for the prostate, bladder, and anterior rectal wall are shown in Fig. 5 a-c.  The overall effect of ERB interventions was a reduction in voxel-wise displacement for all organs under consideration.

To understand the dosimetric effects that interventions have on an entire course of treatment, a dose-volume histogram of the overall treatment effect is shown in Fig. 6.   In fractions where no intervention occurred (n=11), the doses based on the CBCT for that fraction were used for both pre- and post-intervention.  In other words, this figure shows the actual doses delivered alongside the doses that would have been delivered in the absence of interventions in 24 of 35 fractions.  In all three organs under consideration, intervention resulted in a dose distribution that more closely resembled the treatment plan.

 \begin{figure*}
 \includegraphics{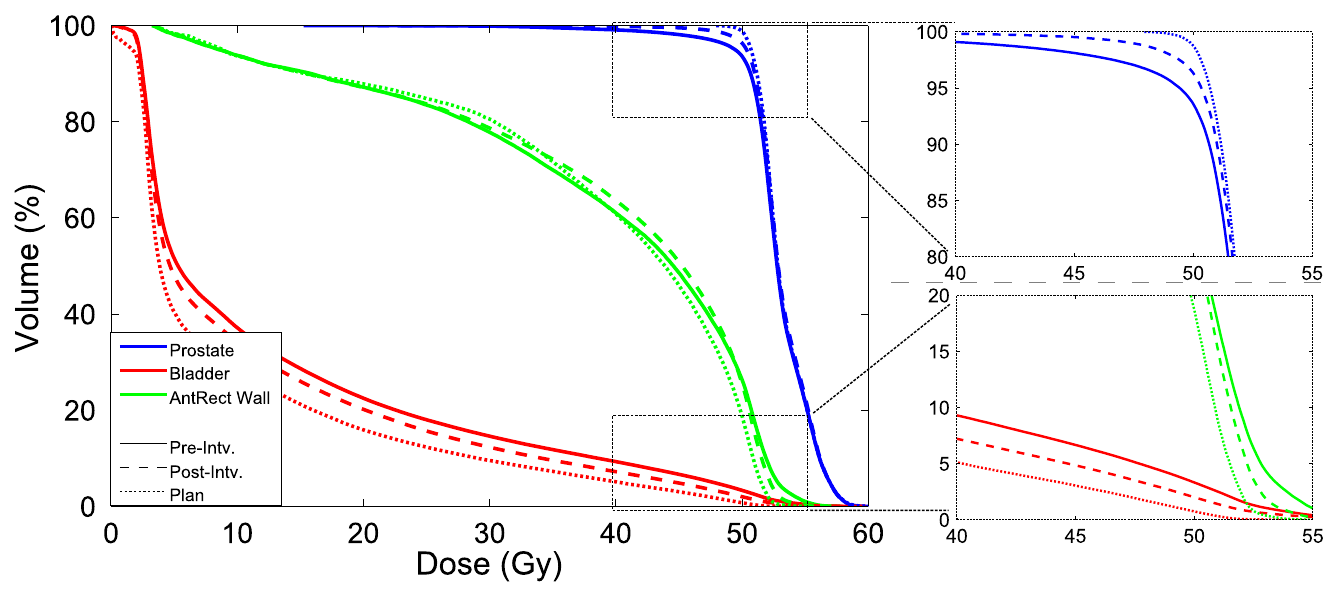}%
 \caption{Cumulative DVH for pre-/post-intervention and planned dose based on all 35 treated fractions.  In fractions where no intervention occurred (n=11), the doses based on the CBCT for that fraction were used for both pre- and post-intervention.  In other words, “Post-Intv” represents the total doses delivered to the patients, and “Pre-Intv” represents the total doses that would have been delivered, had no interventions occurred.}%
 \end{figure*}

\subsection{Effect of Intervention on Prostate Shape}
Figs. 4 and 5a show the displacement in the shape of the prostate between the treatment position and the plan position.  This is useful for comparing the direct effects of ERB interventions on delivered dose; however, it does not give an accurate picture of the true deformation experienced by the prostate because there are two conflicting goals in the alignment of the patient during treatment: PTV coverage and OAR sparing.  Thus, the treatment position may not reflect a true rigid alignment between the planning and treatment prostate shapes.

Fig. 5d shows the cumulative effect of ERB interventions on deformations in the shape of the prostate.  As shown in Fig. 5d, the number of voxels experiencing a deformation in excess of roughly 5 mm decreased, and the number of voxels deformed by less than 5 mm increased.  Additionally, Table 2 shows the effect of ERB interventions on a number of parameters which describe the shape of the prostate.  The most significant improvements in prostate shape were a decrease in AP deformation, a decrease in the tilt of the prostate in the sagittal plane, and an increase in the Dice coefficient between the planning and treatment shapes.  P-values are shown for a paired t-test of the hypothesis of equal means.

\begin{table}[htbp]
  \centering
  \caption{Effect of manual ERB adjustment on prostate shape (n=24)}
    \begin{tabular}{rccccc}
    \hline
    \textbf{} & \multicolumn{2}{c}{Pre-Intervention} & \multicolumn{2}{c}{Post-Intervention} & p \\
    \hline
    \multicolumn{1}{c}{AP def. (mm)} & 3.4   & \enm{\pm}1.1  & 2.8   & \enm{\pm}1.1  & 0.02 \\
    \multicolumn{1}{c}{LR def. (mm)} & 2.6   & \enm{\pm}0.7  & 2.3   & \enm{\pm}0.7  & 0.11 \\
    \multicolumn{1}{c}{SI def. (mm)} & 0.4   & \enm{\pm}0.3  & 0.4   & \enm{\pm}0.3  & 0.73 \\
    \multicolumn{1}{c}{Sagittal Tilt (\enm{^\circ})} & 15    & \enm{\pm}8.2  & 10    & \enm{\pm}6.0  & 0.09 \\
    \multicolumn{1}{c}{Dice Coef.} & 0.76  & \enm{\pm}0.06 & 0.8   & \enm{\pm}0.04 & 0.01 \\
    \hline
    \end{tabular}%
	\enm{} Abbreviations: AP=anterior-posterior, LR=left-right, SI=superior-inferior, def=deformation
  \label{tab:addlabel}%
\end{table}%

\section{Discussion}

A previous study\super{31} found that interventions in rectal and bladder filling were not necessary for patients undergoing post-prostatectomy radiation therapy with conventional fractionation and daily CBCT localization.  However, one expects the analogous scenario of ERB intervention in SBRT to be more significant for several reasons: namely, the increased frequency of intervention (69\% vs. 16\%), higher doses per fraction (10 Gy vs. 2 Gy), and tighter treatment margins (3 mm vs. 5-10 mm).  Using an ERB can be dosimetrically advantageous due to a reduction of intrafraction motion of the prostate\super{16-18} and lower dose to the posterior rectal wall\super{24-26}.  However, ERB use extends the time required for setup/treatment, can cause patient discomfort, and has the possibility to introduce localization errors due to improper positioning and/or inflation.  The results of our study show that ERB position can have a significant impact on the dosimetry of each fraction, and that manual interventions can bring about an increase in prostate D95 and coverage.  It should be noted that the effects presented here were seen using an ERB with 60 cm\super{3} volume. As a larger/smaller ERB would exert different forces on the prostate, these results may differ for other sizes of ERB in clinical use.

 \begin{figure}[h]
 \includegraphics{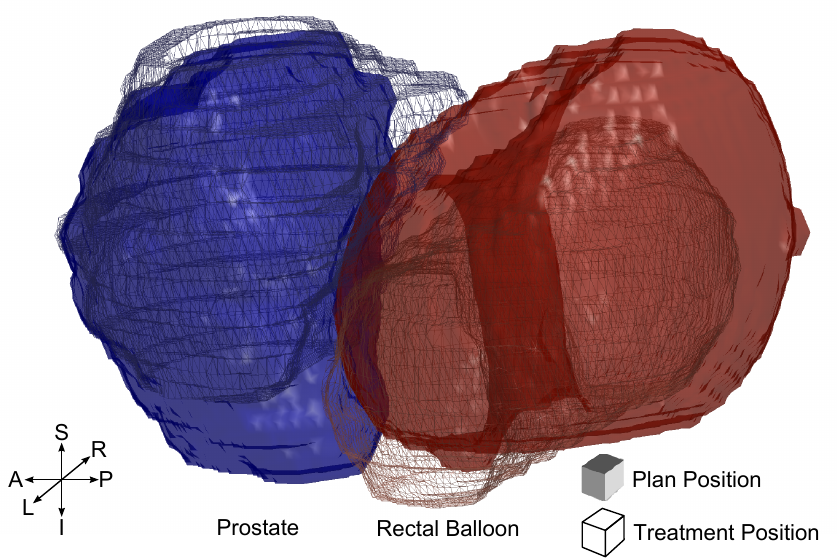}%
 \caption{Effect of rectal balloon (ERB) position on prostate deformation in one sample treatment fraction.  A patient is simulated with an ERB originally at position (A) in the plan CT.  During treatment, the ERB is placed at position (B), and the anterior edge of the ERB is displaced towards the base of the prostate.  This causes the prostate, originally simulated in position (C), to be displaced/deformed superiorly to position (D).}%
 \end{figure}

\subsection{Prostate Metrics}
We have shown here that interventions in ERB position lead to improved dosimetric outcomes for the treatment of prostate cancer.  This is evidenced by the post-intervention decreases in dose difference and surface displacement (Fig. 4), post-intervention increases in prostate D95 and coverage (Table 1), decreases in displacement and deformation of the prostate (Fig. 5a and d), and improved overall DVH characteristics (Fig. 6).   While the prostate dosimetric outcomes are higher in the post-intervention case, it is not certain whether or not the differences in the population-averaged metrics are clinically significant.  However, it is apparent from Fig. 4 that the effects of interventions can be quite large on the individual level.  Taken together, these results suggest that ERB interventions are necessary to ensure the delivery of dose distributions as planned.

This has several implications for radiotherapy of the prostate using SBRT.  First, it implies that two CBCT scans are generally needed for adequate target localization: one after insertion of the ERB and one after adjustment (if necessary).  Second, it supports the idea that careful review of the patient anatomy and target localization are crucial for target coverage.

\subsection{ERB Position}
We found that ERB position interventions lead to an overall decrease in the amount of deformation of the prostate compared to the planning position (Fig. 5d and Table 2).  Fig. 7 demonstrates one possible mechanism for these effects.  This plot shows the structures from the pCT and CBCT of a particular pre-intervention fraction.  In this setup, the ERB centroid is lower than planned as a result of insufficient insertion depth.  This manifests as an inferior shift in the point of contact between the anterior rectal wall and the ERB.  As a result, the prostate is deformed in the AP and superior-inferior (SI) directions, and tilted in the sagittal plane.  Since this shape differs from the planning prostate, it is difficult to cover the prostate with the existing fields, which are shaped by the MLCs to conform very precisely to the pCT prostate shape.  In other words, no combination of rigid translations will allow the conformal treatment fields to cover a deformed prostate, and coverage suffers as a result.

One possible solution to this problem would be to generate an updated treatment plan based on the patient’s daily anatomy (Adaptive Radiotherapy).  However, this carries its own set of challenges, especially in terms of the reduced quality of CBCT, the workload required to generate a plan in real time, and the ability to verify the treatment before delivery.

\subsection{Anterior Rectal Wall and Bladder}
There was no significant reduction or improvement in dose metrics for the anterior rectal wall or bladder.  However, the overall effect of ERB interventions was to reduce deformations in these structures between the planning and treatment position, resulting in delivered dose distributions that more closely resembled the treatment plan.  Improper ERB position can push the prostate out of the field, leading to a decrease in coverage; however, if the ERB were placed incorrectly it could cause either an increase or decrease in dose to the rectum and bladder.  It was seen that there was no change in mean value, as these changes negate each other when calculating the dose metrics for anterior rectal wall or bladder. However, these post-intervention values conform better to the treatment plan (Fig. 5).  

\section{Conclusions}

With the use of ERBs during prostate SBRT, interventions in ERB position are frequently necessary to ensure the delivery of the dose distribution as planned. We found that interventions lead to significant post-intervention increases in prostate D95 and coverage, decreases in displacement at the time of treatment of the prostate, and improved post-intervention DVH characteristics for PTV, anterior rectal wall, and bladder.  We also found that ERB adjustments lead to decreases in deformation of the prostate in the AP direction, decreases in the tilt of the prostate from the planning position, and an increase in the similarity in shape (Dice coefficient) between the planning and treatment prostate shapes.  

% \begin{table}
% \end{tabular}
% \end{table}

%\begin{acknowledgments}
\section*{Acknowledgements}
The authors would like to thank Dr. Teisha Rowland for her help in preparing the manuscript for publication.  This work was partially supported by DOD Grant PC061629.
%\end{acknowledgments}

\section*{References}
{\small 1.	B. D. Kavanagh, M. Miften and R. A. Rabinovitch, "Advances in Treatment Techniques: Stereotactic Body Radiation Therapy and the Spread of Hypofractionation," The Cancer Journal 17, 177-181 (2011).

2.	S. Jabbari, V. K. Weinberg, T. Kaprealian, I. C. Hsu, L. Ma, C. Chuang, M. Descovich, S. Shiao, K. Shinohara, M. Roach Iii and A. R. Gottschalk, "Stereotactic Body Radiotherapy as Monotherapy or Post–External Beam Radiotherapy Boost for Prostate Cancer: Technique, Early Toxicity, and PSA Response," International Journal of Radiation Oncology*Biology*Physics (2011).

3.	A. Katz, M. Santoro, R. Ashley, F. Diblasio and M. Witten, "Stereotactic body radiotherapy for organ-confined prostate cancer," BMC Urology 10, 1-10 (2010).

4.	D. E. Freeman and C. R. King, "Stereotactic body radiotherapy for low-risk prostate cancer: five-year outcomes," Radiation Oncology 6, 3 (2011).

5.	C. R. King, J. D. Brooks, H. Gill, T. Pawlicki, C. Cotrutz and J. C. Presti Jr, "Stereotactic Body Radiotherapy for Localized Prostate Cancer: Interim Results of a Prospective Phase II Clinical Trial," International Journal of Radiation Oncology*Biology*Physics 73, 1043-1048 (2009).

6.	T. P. Boike, Y. Lotan, L. C. Cho, J. Brindle, P. DeRose, X.-J. Xie, J. Yan, R. Foster, D. Pistenmaa, A. Perkins, S. Cooley and R. Timmerman, "Phase I Dose-Escalation Study of Stereotactic Body Radiation Therapy for Low- and Intermediate-Risk Prostate Cancer," Journal of Clinical Oncology 29, 2020-2026 (2011).

7.	B. L. Madsen, R. A. Hsi, H. T. Pham, J. F. Fowler, L. Esagui and J. Corman, "Stereotactic hypofractionated accurate radiotherapy of the prostate (SHARP), 33.5 Gy in five fractions for localized disease: First clinical trial results," International Journal of Radiation Oncology*Biology*Physics 67, 1099-1105 (2007).

8.	D. J. Brenner and E. J. Hall, "Fractionation and protraction for radiotherapy of prostate carcinoma," International Journal of Radiation Oncology*Biology*Physics 43, 1095-1101 (1999).

9.	S. G. Williams, J. M. G. Taylor, N. Liu, Y. Tra, G. M. Duchesne, L. L. Kestin, A. Martinez, G. R. Pratt and H. Sandler, "Use of Individual Fraction Size Data from 3756 Patients to Directly Determine the α/β Ratio of Prostate Cancer," International Journal of Radiation Oncology*Biology*Physics 68, 24-33 (2007).

10.	C. Fiorino, F. Foppiano, P. Franzone, S. Broggi, P. Castellone, M. Marcenaro, R. Calandrino and G. Sanguineti, "Rectal and bladder motion during conformal radiotherapy after radical prostatectomy," Radiotherapy and Oncology 74, 187-195 (2005).

11.	J. A. Antolak, I. I. Rosen, C. H. Childress, G. K. Zagars and A. Pollack, "Prostate target volume variations during a course of radiotherapy," International Journal of Radiation Oncology*Biology*Physics 42, 661-672 (1998).

12.	J. M. Crook, Y. Raymond, D. Salhani, H. Yang and B. Esche, "Prostate motion during standard radiotherapy as assessed by fiducial markers," Radiotherapy and Oncology 37, 35-42 (1995).

13.	J. C. Roeske, J. D. Forman, C. F. Mesina, T. He, C. A. Pelizzari, E. Fontenla, S. Vijayakumar and G. T. Chen, "Evaluation of changes in the size and location of the prostate, seminal vesicles, bladder, and rectum during a course of external beam radiation therapy," International journal of radiation oncology, biology, physics 33, 1321-1329 (1995).

14.	M. van Herk, A. Bruce, A. P. Kroes, T. Shouman, A. Touw and J. V. Lebesque, "Quantification of organ motion during conformal radiotherapy of the prostate by three dimensional image registration," International Journal of Radiation Oncology, Biology, Physics 33, 1311-1320 (1995).

15.	E. Huang, L. Dong, A. Chandra, D. A. Kuban, I. I. Rosen, A. Evans and A. Pollack, "Intrafraction prostate motion during IMRT for prostate cancer," International Journal of Radiation Oncology*Biology*Physics 53, 261-268 (2002).

16.	C. Vargas, A. I. Saito, W. C. Hsi, D. Indelicato, A. Falchook, Q. Zengm, K. Oliver, S. Keole and J. Dempsey, "Cine-Magnetic Resonance Imaging Assessment of Intrafraction Motion for Prostate Cancer Patients Supine or Prone With and Without a Rectal Balloon," American Journal of Clinical Oncology 33, 11-16 (2010).

17.	K. K. Wang, N. Vapiwala, C. Deville, J. P. Plastaras, R. Scheuermann, H. Lin, V. Bar Ad, Z. Tochner and S. Both, "A Study to Quantify the Effectiveness of Daily Endorectal Balloon for Prostate Intrafraction Motion Management," International Journal of Radiation Oncology*Biology*Physics (2011). http://dx.doi.org/10.1016/j.ijrobp.2011.07.038

18.	R. J. Smeenk, R. J. Louwe, K. M. Langen, A. P. Shah, P. A. Kupelian, E. N. van Lin and J. H. Kaanders, "An Endorectal Balloon Reduces Intrafraction Prostate Motion During Radiotherapy," International Journal of Radiation Oncology*Biology*Physics (2011), http://dx.doi.org/10.1016/j.ijrobp.2011.07.028.

19.	J. M. Hensel, C. Ménard, P. W. M. Chung, M. F. Milosevic, A. Kirilova, J. L. Moseley, M. A. Haider and K. K. Brock, "Development of Multiorgan Finite Element-Based Prostate Deformation Model Enabling Registration of Endorectal Coil Magnetic Resonance Imaging for Radiotherapy Planning," International Journal of Radiation Oncology*Biology*Physics 68, 1522-1528 (2007).

20.	J. Lian, L. Xing, S. Hunjan, C. Dumoulin, J. Levin, A. Lo, R. Watkins, K. Rohling, R. Giaquinto, D. Kim, D. Spielman and B. Daniel, "Mapping of the prostate in endorectal coil-based MRI/MRSI and CT: A deformable registration and validation study," Medical Physics 31, 3087-3094 (2004).

21.	M. Zaider, M. J. Zelefsky, E. K. Lee, K. L. Zakian, H. I. Amols, J. Dyke, G. Cohen, Y.-C. Hu, A. K. Endi, C.-S. Chui and J. A. Koutcher, "Treatment planning for prostate implants using magnetic-resonance spectroscopy imaging," International Journal of Radiation Oncology*Biology*Physics 47, 1085-1096 (2000).

22.	T. Mizowaki, G. a. N. Cohen, A. Y. C. Fung and M. Zaider, "Towards integrating functional imaging in the treatment of prostate cancer with radiation: the registration of the MR spectroscopy imaging to ultrasound/CT images and its implementation in treatment planning," International Journal of Radiation Oncology*Biology*Physics 54, 1558-1564 (2002).

23.	S. W. Heijmink, T. W. Scheenen, E. N. van Lin, A. G. Visser, L. A. Kiemeney, J. A. Witjes and J. O. Barentsz, "Changes in Prostate Shape and Volume and Their Implications for Radiotherapy After Introduction of Endorectal Balloon as Determined by MRI at 3T," International Journal of Radiation Oncology*Biology*Physics 73, 1446-1453 (2009).

24.	S. Wachter, N. Gerstner, D. Dorner, G. Goldner, A. Colotto, A. Wambersie and R. Pötter, "The influence of a rectal balloon tube as internal immobilization device on variations of volumes and dose-volume histograms during treatment course of conformal radiotherapy for prostate cancer," International Journal of Radiation Oncology*Biology*Physics 52, 91-100 (2002).

25.	I. F. Ciernik, B. G. Baumert, P. Egli, C. Glanzmann and U. M. Lütolf, "On-line correction of beam portals in the treatment of prostate cancer using an endorectal balloon device," Radiotherapy and Oncology 65, 39-45 (2002).

26.	R. J. Smeenk, E. N. van Lin, P. van Kollenburg, G. M. McColl, M. Kunze-Busch and J. H. Kaanders, "Endorectal balloon reduces anorectal doses in post-prostatectomy intensity-modulated radiotherapy," Radiotherapy and Oncology 101, 465-470 (2011).

27.	B. S. Teh, J. E. McGary, L. Dong, W.-Y. Mai, L. S. Carpenter, H. H. Lu, J. K. Chiu, S. Y. Woo, W. H. Grant and E. B. Butler, "The Use of Rectal Balloon During the Delivery of Intensity Modulated Radiotherapy (IMRT) for Prostate Cancer: More Than Just a Prostate Gland Immobilization Device?," The Cancer Journal 8, 476-483 (2002).

28.	B. Emami, J. Lyman, A. Brown, L. Cola, M. Goitein, J. E. Munzenrider, B. Shank, L. J. Solin and M. Wesson, "Tolerance of normal tissue to therapeutic irradiation," International Journal of Radiation Oncology*Biology*Physics 21, 109-122 (1991).

29.	C. Fiorino, G. Sanguineti, C. Cozzarini, G. Fellin, F. Foppiano, L. Menegotti, A. Piazzolla, V. Vavassori and R. Valdagni, "Rectal dose–volume constraints in high-dose radiotherapy of localized prostate cancer," International Journal of Radiation Oncology*Biology*Physics 57, 953-962 (2003).

30.	L. J. Boersma, M. van den Brink, A. M. Bruce, T. Shouman, L. Gras, A. te Velde and J. V. Lebesque, "Estimation of the Incidence of Late Bladder and Rectum Complications After High-Dose (70–78 Gy) Conformal Radiotherapy for Prostate Cancer, Using Dose–Volume Histograms," International Journal of Radiation Oncology*Biology*Physics 41, 83-92 (1998).

31.	Q. Diot, C. Olsen, B. Kavanagh, D. Raben and M. Miften, "Dosimetric Effect of Online Image-Guided Anatomical Interventions for Postprostatectomy Cancer Patients," International Journal of Radiation Oncology*Biology*Physics 79, 623-632 (2011).

32.	Simmons Cancer Center; University of Texas Southwestern Medical Center. “Stereotactic Body Radiation Therapy in Treating Patients With Prostate Cancer”. In: ClinicalTrials.gov [Internet]. Bethesda (MD): National Library of Medicine (US). 2000- [cited 2011 Nov 30]. Available from: http://clinicaltrials.gov/ct/show/NCT00547339 NLM Identifier: NCT00547339.

33.	G. J. Meijer, M. van den Brink, M. S. Hoogeman, J. Meinders and J. V. Lebesque, "Dose–wall histograms and normalized dose–surface histograms for the rectum: a new method to analyze the dose distribution over the rectum in conformal radiotherapy," International Journal of Radiation Oncology*Biology*Physics 45, 1073-1080 (1999).

34.	M. H. Davis, A. Khotanzad, D. P. Flamig and S. E. Harms, "A physics-based coordinate transformation for 3-D image matching," IEEE Trans Med Imag 16, 317-328 (1997).

35.	A. Richter, Q. Hu, D. Steglich, K. Baier, J. Wilbert, M. Guckenberger and M. Flentje, "Investigation of the usability of conebeam CT data sets for dose calculation," Radiation Oncology 3, 42 (2008).

36.	L. R. Dice, "Measures of the amount of ecologic association between species," Ecology 26, 297-302 (1945).}

%\bibliography{your-bib-file}

\end{document}